# LTE-ADVANCED SIMULATION FRAMEWORK BASED ON OMNeT++

Muhsen Hammoud, Assoc. Prof. Dr. Abid Abdelouahab
University Tenaga Nasional
Email: mohsen_hammoud@yahoo.com, abid@uniten.edu.my

*Abstract* Long Term Evolution-Advanced (LTE-Advanced) is the most recent mobile telecommunication technology proposed by 3GPP. LTE-Advanced is applied in some countries, but still in development and testing phase, because of that, a simulation model is needed to test this technology. This paper introduce a LTE-Advanced simulation framework based on OMNeT++ IDE (open source). The proposed framework test results showed that it's working properly and ready to be used. This framework can be considered as the beginning of the LTE-Advanced simulation model based on OMNeT++.

I. INTRODUCTION

Future mobile communication systems expectations and requirements continue to grow and evolve[1]. Therefore many organizations that are specialized in specifications have been working on a competitive mobile communication system. 3GPP as a leading organization in mobile communication specification had considered Long Term Evolution-Advanced (LTE-Advanced) mobile communication system as a competitive system in the future. LTE is an evolved radio interface merged with the use of Orthogonal Frequency-Division Multiple Access (OFDMA) and Single Carrier Frequency Domain Multiple Access (SC-FDMA).

For such a new technology, testing in real world would be of high cost, besides the time consumed to complete the tests. As a solution for this problem, modeling and simulations are used.

The LTE-Advanced simulation model framework proposed in this paper contains all protocols involved in LTE-Advanced protocol stack. These protocols are like black boxes in this framework, and in the next step, all protocols will be analyzed and implemented.

The main aim of creating a simulation model is to answer these questions: What to include in the simulation? And how to rewrite the algorithms of the protocols in such a way that gives the needed results as in the specifications documents, and in the same time to be suitable for simulation model?

The rest of the paper is organized as follows: Section II describes LTE and LTE-Advanced. Sections III presents the architecture of LTE-Advanced network and its components. Section IV describes the development environment. Section V contains explanation of framework implementation. Section VI contains the testing methodology and the results. Finally, Section VII draws the conclusion.

II. LTE AND LTE-ADVANCED

First part will describe LTE networks, second part will describe LTE-Advanced.

A. *LTE*

LTE is a new set of standards for high-speed mobile communication for data terminals and mobile phones. LTE is based on GSM/EDGE and UMTS/HSPA technologies. The main aim of LTE was to guarantee competitiveness for a 10 years' time frame. The requirements of LTE were refined till June 2005 when it was finalized.

A summarize for these requirements will be as follows:
- Reduced delays, in terms of both connection establishment and transmission latency.
- Increased user data rates.
- Increased cell-edge bit-rate, for uniformity of service provision.
- Reduced cost per bit, implying improved spectral efficiency.
- Greater flexibility of spectrum usage, in both new and pre-existing bands.
- Simplified network architecture.
- Seamless mobility, including between different radio-access technologies.
- Reasonable power consumption for the mobile terminal.

The next phase of LTE which is LTE-Advanced are guided by the requirements determined by the Next Generation Mobile Networks (NGMN) union of network operators[2].

B. *LTE-Advanced*

The project of LTE-Advanced started in October 2009 by 3GPP, when 3GPP submitted this project to the International Telecommunication Union (ITU) as a suggested candidate IMT-Advanced technology, the specifications of this project are said to be available in 2011 by Release-10[3].

The intent of LTE-Advanced was mainly to improve LTE radio access, the improvement is in terms of system performance and capabilities in comparison with the current mobile systems. The main goal was to make sure that LTE operates at or even beyond the requirements of IMT-Advanced as defined by ITU-R[4][5][6].

### III. LTE NETWORK ARCHITECTURE

LTE basic system architecture consists of 4 parts: the User Equipment (UE) domain, Evolved-Universal Terrestrial Radio Access Network (E-UTRAN) domain, Evolved Packet Core (EPC) domain, and the Services domain. The Evolved Packet System (EPS) consists of the first three components (UE, E-UTRAN, and EPC), EPS is IP based connectivity[7]. Figure 1 illustrates these 4 domains.

The User Equipment (UE) domain consists of mobile users with their User Equipment.

The E-UTRAN domain consists of evolved NodeBs (eNodeBs), which are the base stations distributed all over the network coverage area. The main aim of eNodeB is to forward data between UE and EPC, this means between radio connection and IP connection. eNodeB also have many functions:

- Coding/decoding of user plan data.
- Header compression/decompression for IP packets.
- Some control plan functionalities.
- Radio Resource Management (RRM).
- Mobile management (MM).

MME represents the main control element in the EPC domain. Many functions for MME:

- Security and authentication.
- Subscriber's profile management.
- Mobility management.
- Service connectivity.

S-GW main role is user plan tunneling management and switching. IP packets is the format of user data, these packets are encapsulated in GTP packets, this process is GTP tunneling. The aim of this process is to control the tunnels to eNodeB. When a UE moves from one eNodeB to another, S-GW switches the tunnel of this UE to the new eNodeB under the control of MME. Tunneling process includes relaying user data between eNodeB and P-GW, besides monitoring the data, also it's able to perform lawful interception.

P-GW represents the edge router between EPC and service domain. P-GW is called the default gateway because it is the IP attachment point of UE. P-GW have many functions:

- Allocate IP address to the UE.
- DHCP functionality.
- IP data flow mapping into GTP tunnels.

Home Subscription Server (HSS) is a repository for all user data, like registering the location of a user in a visited network. It can be said that HSS is like a database server stores all information about the user and services that are applicable to the user.

The last domain is the Service Domain, it consists of many kinds of services like: IP Multimedia Sub-system (IMS) and video streaming services among other service.

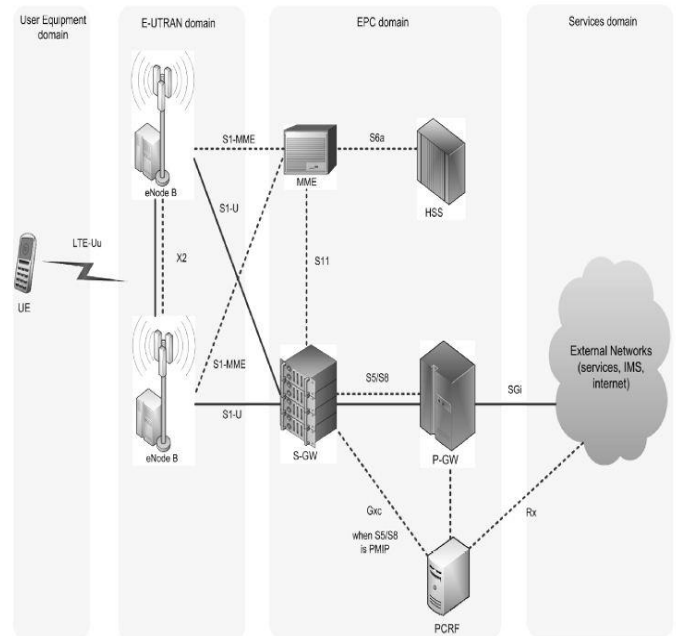

Figure 1. Basic system architecture of the LTE-Advanced network[7]

### IV. DEVELOPMENT ENVIRONMENT

Objective Modular Network Testbed in C++ (OMNet++) is used to develop the framework. OMNeT++ framework has been developed by András Varga, at the Technical University of Budapest, Department of Telecommunications[8].

OMNeT++ is a discrete event object-oriented network simulation framework. It's architecture is generic, this architecture allows it to be used in various problem domains like:

- Modeling of wired and wireless communication networks.
- Protocol modeling.
- Modeling of queuing networks.
- Modeling of multiprocessors and other distributed hardware systems.
- Validation of hardware architecture.
- Evaluating performance aspects of complex software systems.
- Modeling and simulation of any system with discrete event approach.

OMNeT++ provides a suite infrastructure and a variety of tools to write simulations, it's not a simulator itself..

Since the main aim of this research is to create a simulation model for LTE protocol stack, and the current models have many defects, OMNeT++ is the chosen IDE to implement the proposed LTE framework.

### V. FRAMEWORK IMPLEMENTATION

The framework consists of four elements: UE, eNB, S-GW/MME, and PDN-GW. The LTE protocol stack is between UE and eNB except for NAS protocol which is between UE and MME. The S-GW and MME is compound in the same component since these two components of the network don't affect the performance of the LTE in a direct way. Inside each compound module there are number of simple modules represent the protocol stack.

Generally, the network have one PDN-GW, one S-GW/MME, many eNBs, and many UEs, to enable the user to build a network like the one in figure 2.

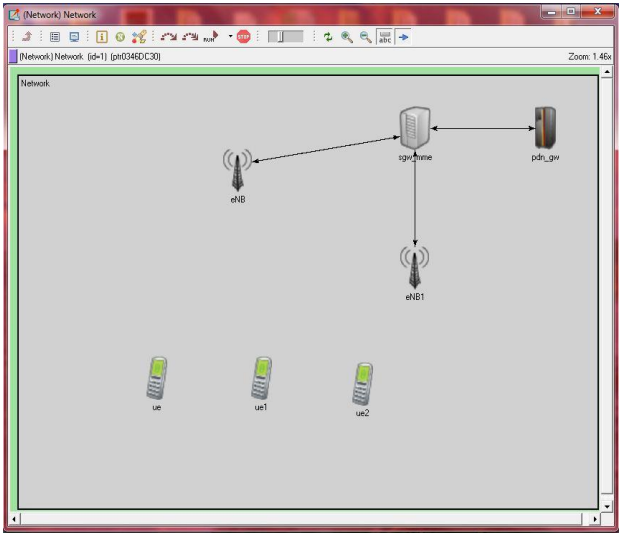

Figure 2. General Network Description

The User Equipment(UE) implementation will be described briefly, the other componants implementation is similar to UE implementation. User equipment (UE) consists of 6 layers or protocols: NAS, RRC, PDCP, RLC, MAC, and PHY. Each protocol is implemented as a simple module, then these modules are combined together to form the UE device, figure 3.

NAS is the highest part of the protocol stack, it has only one channel to the lower layer which is RRC, this channel is two ways channel. When this simple module receives packets or messages from RRC, it just drops it since it's not implemented in this research.

RRC is connected to both NAS and PDCP with a bidirectional channels. This simple module has a simple code to pass messages only and flag them with its name.

PDCP, RLC are explained later in this chapter.

MAC is like RRC just connected to RLC and PHY by bidirectional channel. There is a simple code to control this simple module, this code is just for passing the incoming messages either to upper layer or lower layer and flag the message by a its name.

PHY is the last layer in the UE, it either receives messages from MAC and sends them to the eNB by using the function **sendDirect**, or receives messages from the eNB. **sendDirect** function sends the message or the packet through air directly to the radio interface of the target eNB which is already defined in the UE. PHY is connected to the radio interface of the UE, this radio interface receives incoming messages from eNB which the UE is attached to and pass them to the physical layer to be processed properly.

The code that passes messages through the simple module is:

```
int arrivedGate = msg->getArrivalGateId();
if(arrivedGate == gate("inFromUpperLayer")->getId())
{
    if(msg->isPacket())
    {
        cPacket *rec =
    check_and_cast<cPacket *>(msg);
        rec->setName("MACPck");
        send(rec, "outToLowerLayer");
    }
    else
    {
        send(msg, "outToLowerLayer");
    }
}
else
{
    if(msg->isPacket())
    {
        cPacket *rec =
    check_and_cast<cPacket *>(msg);
        rec->setName("RLCPck");
        send(rec, "outToUpperLayer");
    }
    else
    {
        send(msg, "outToUpperLayer");
    }
}
```

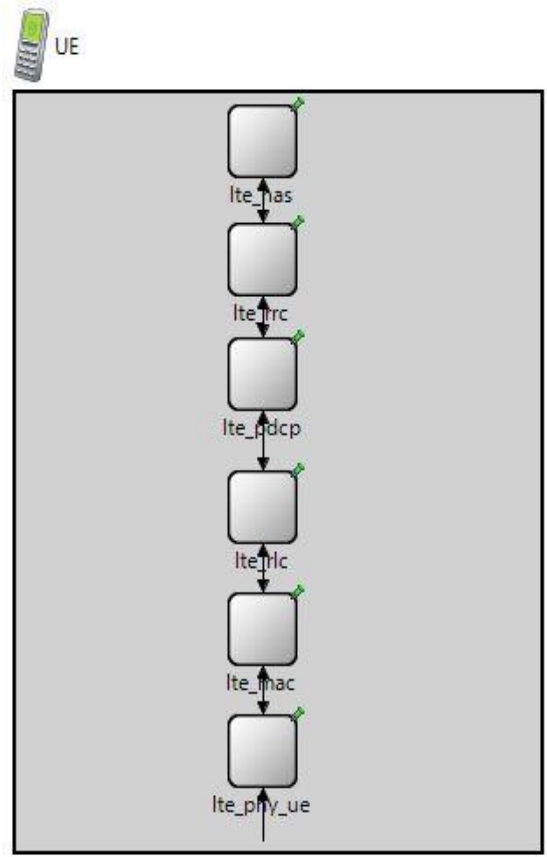

Figure 3. UE Implementation

## VI. FRAMEWORK TESTING RESULTS

To test the framework, a simple module called Generator is created and attached to NAS layer in UE, the function of this module is to generate messages repeatedly using a fixed time period, the time used in this test is 0.01 seconds. The test is done on the network in the figure 4. The aim of this test is to make sure that the messages go through each module, and to make sure that the channels are working in the right way. Figure 5 shows the results of this test. The format of the results is like the following lines:

** Event #1  T=0  Network.ue.lte_nas (lte_nas, id=18), on `NASMsg' (cMessage, id=2)

** Event #2  T=0  Network.ue.lte_rrc (lte_rrc, id=17), on `RRCMsg' (cMessage, id=2)

Event #1 is created at time 0, the message passed NAS protocol in UE, the message name is NASMsg, then when it goes out to RRC protocol, it become RRCMsg at Event #2. The rest of the results is read in the same way. From the results, the message generated by generator module in UE is going through the UE layers first, then it is sent through air to the eNB, the message goes inside the layers of eNB then goes out to the S-GW/MME and travels through its layers to reach the out gate towards PDN_GW, in PDN_GW the message goes through the layers to be sent again in the opposite direction back in the same way till it reaches the generator module in UE, where it's discarded, and a new message is created again to go in the same way till the user stops the simulator.

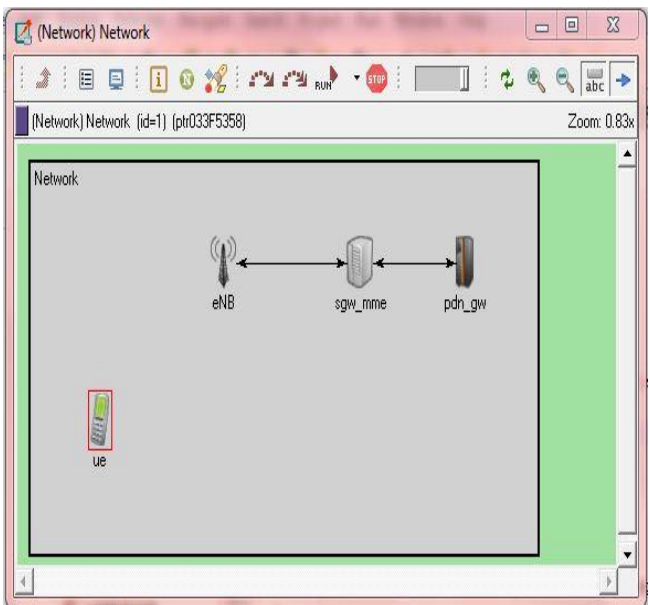

Figure 4. Tested Network Architecture

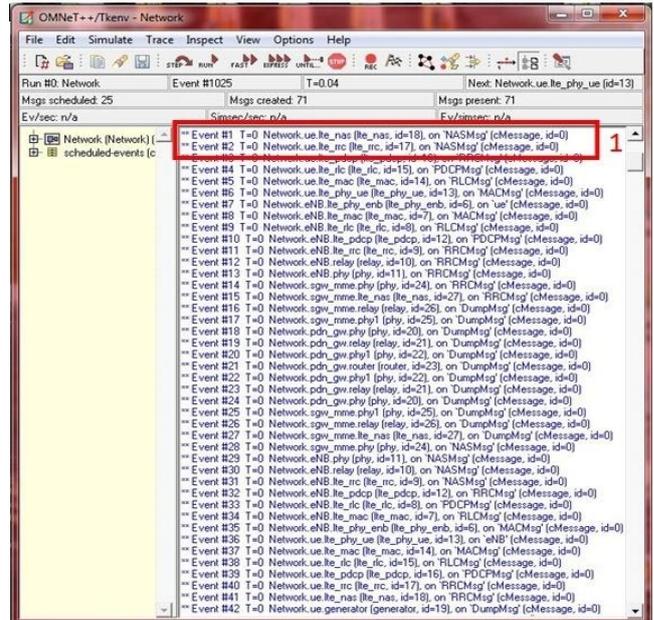

Figure 5. Framework Testing Results

## VII. CONCLUSION

The testing results of the framework showed that it's working properly, all involved modules pass messages and packets in a proper way, and in the right format.